\documentclass{rnaastex}

\begin{document}

\title{Limits on Ionized Gas in M81's Globular Clusters}

%% Note that the corresponding author command and emails has to come
%% before everything else. Also place all the emails in the \email
%% command instead of using multiple \email calls.
\correspondingauthor{J. M. Wrobel}
\email{jwrobel@nrao.edu}

%% The \author command can take an optional ORCID.
\author[0000-0001-9720-7398]{J. M. Wrobel}
\affiliation{National Radio Astronomy Observatory, P.O. Box O,
  Socorro, NM 87801, USA}

\author[0000-0001-8348-2671]{K. E. Johnson}
\affiliation{Department of Astronomy, University of Virginia, 530
  McCormick Road, Charlottesville, VA 22904, USA}

%% Note that RNAAS manuscripts DO NOT have abstracts.


%% Start the main body of the article. If no sections in the 
%% research note leave the \section call blank to make the title.
\section{} 

Observational constraints on the gas in globular star clusters (GCs)
are key to understanding GC evolution \citep[][and references
therein]{rob88,van09}.  As the stars in GCs evolve, they shed gas into
the potential wells of the GCs.  Rough estimates suggest that about
100 to 1000~$M_\odot$ of gas could be accumulated before being purged
as the GCs cross the Galactic disk.  Searches have failed to detect
these amounts of gas in Galactic GCs.  Still, observations indicate
about 0.1~$M_\odot$ of ionized gas in the central few parsecs of 47
Tuc and M15 \citep{pfa01,fre01} and (tentatively) about 0.3 $M_\odot$
of neutral gas in M15 \citep{van06}.  Numerical studies of gas removal
between disk crossings suggest a diversity of gas properties among
Galactic GCs \citep{pri11,pep13,mcd15,pep16}.

Measurements encompassing the bulk of a GC should offer the most
robust constraints on its gas content.  This can be problematic for
Galactic GCs because they subtend arcminutes.  For example, some gas
inventories require extrapolations, while others could be affected by
discrete sources in front of, within, or behind the GCs, or by the
Galactic interstellar medium toward the GCs
\citep[e.g.,][]{hil73,kna96,van09}.  These concerns are mitigated for
extragalactic GCs.  Their typical half-light radii of 2-3 pc
\citep{bro06} mean they subtend less than an arcsecond beyond the
Local Group, making it easier to encompass the bulk of a GC.  Also,
the entire GC system of a galaxy can be studied, important given the
aforementioned diversity of gas properties.

Here, we re-purpose data from our radio search for the radiative
signatures of accretion onto intermediate-mass black holes in probable
GCs in M81, a spiral galaxy at a distance of 3.63 Mpc \citep{wro16}.
Among the 214 probable GCs tabulated by \citet{nan11}, 40\% are
spectroscopically confirmed and 60\% are good GC candidates due to
their colors and sizes.  The 214 probable GCs have a median 2D
half-light radius of 2.63~pc, derived from individual half-light radii
with 40\% uncertainties.  Our radio search used the NRAO Karl G.\
Jansky Very Large Array \citep[VLA;][]{per11} at a wavelength of
5.5~cm and resolution of 1\farcs5 (26.4~pc).  Eight probable GCs fell
outside the search region.  We achieved 3$\sigma$ upper limits of
between $3 \times 4.3$ and $3 \times 51~\mu$Jy~beam$^{-1}$ for
individual GCs and of $3 \times 0.43~\mu$Jy~beam$^{-1}$ for the
weighted-mean image stack of 206 GCs.

We use equation (3) from \citet{kna96} to convert the flux-density
upper limit for each GC and the GC stack to an upper limit on the
ionized gas mass, $M_{\rm HII}$ and $M_{\rm HII}^{\rm stack}$,
respectively.  We assume that the gas is optically thin, isothermal
with $T_e \sim 10^4$ K, of uniform density, and spherically
distributed with a 3D radius that is $\frac{4}{3}$ times the 2D
half-light radius \citep{spi87}.  We adopt the median half-light
radius rather than the poorly-constrained individual radii.  The gas
thus subtends 0\farcs4 arcsec (7~pc), necessitating a 7\% correction
to the flux-density upper limits.  The stellar mass, $M_\star$, for
each GC was taken from \citet{wro16}.  Using their approach, the peak
of the GC luminosity function \citep{nan11} converts to a stellar mass
$M_\star^{\rm peak} \sim 1.6 \times 10^5 M_\odot$.

Figure~1 shows $M_{\rm HII}$ and $M_\star$ for the GCs and locates the
stack's $M_{\rm HII}^{\rm stack}$ at $M_\star^{\rm peak}$.  From
Figure~1 we find: (1) None of the 206 individual GCs in M81 has
ionized gas detected with these radio observations.  A typical
gas-mass limit is $M_{\rm HII} < 550~M_\odot$.  Gas-mass fractions,
$M_{\rm HII} / M_\star$, are below about 0.1 at $M_\star \sim
10^4~M_\odot$ and below about 0.0002 at $M_\star \sim 3.4 \times
10^6~M_\odot$.  (2) From the stack of 206 GCs, the formal gas-mass
limit is about $M_{\rm HII}^{\rm stack} < 150~M_\odot$.  Dividing this
by the stellar mass peak for M81's GCs, we infer a formal gas-mass
fraction below about 0.0009.

These first-look gas constraints are referenced to the median 2D
half-light radius of the GCs.  As such, they provide reasonably robust
constraints on one possible gas phase among the 206 probable GCs in
M81.  These constraints can be improved with longer VLA exposures or
similar exposures with the next-generation VLA \citep{car15}, and with
more accurate half-light radii for the individual GCs.

\begin{figure}[htb]
\plotone{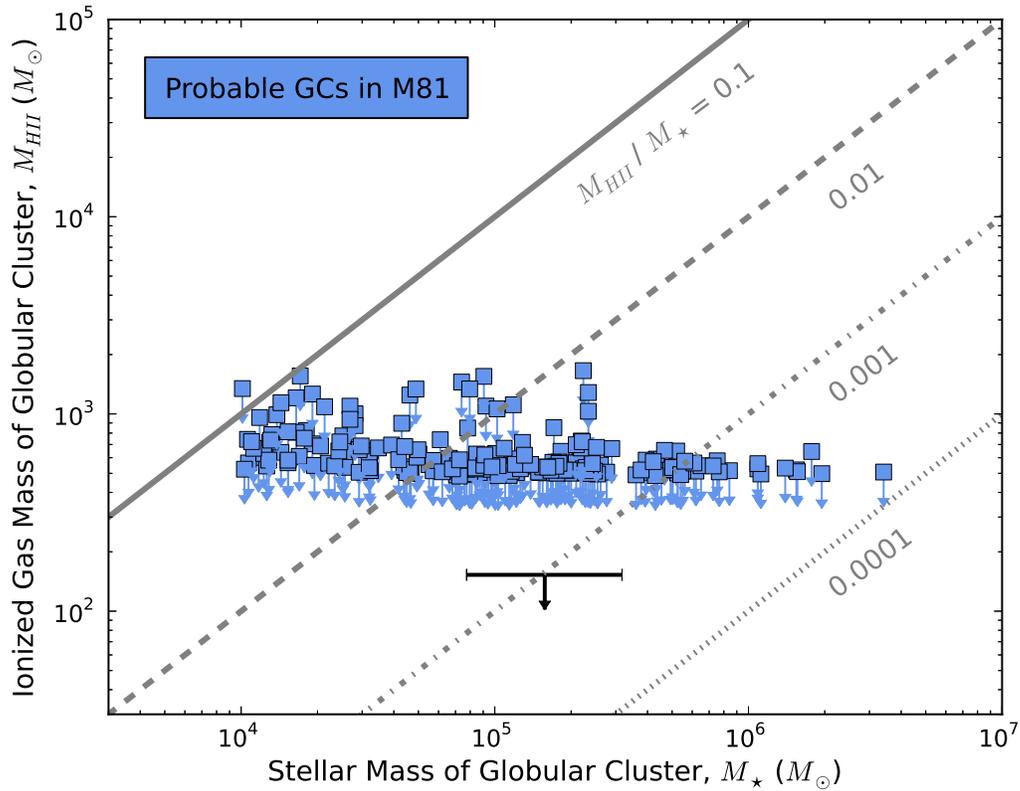}
\caption{Ionized gas mass, $M_{\rm HII}$, and stellar mass, $M_\star$,
  for each probable GC in M81.  The grey diagonal lines of constant
  $M_{\rm HII} / M_\star$ convey ionized-gas-mass fractions.  For the
  stack of 206 probable GCs, $M_{\rm HII}^{\rm stack}$ is plotted as a
  dark line centered at $M_\star^{\rm peak}$.  All gas-mass upper
  limits are at 3$\sigma$. \label{fig:1}}
\end{figure}

\acknowledgments The National Radio Astronomy Observatory (NRAO) is a
facility of the National Science Foundation, operated under
cooperative agreement by Associated Universities, Inc.  K.E.J. is
supported by NSF grant 1413231.

\end{document}